\documentclass[aps,twocolumn,prl,floatfix,superscriptaddress,showpacs,longbilbiography]{revtex4-2}
\usepackage[utf8]{inputenc}

\usepackage{graphicx}
\usepackage{dcolumn}
\usepackage{bm}
\usepackage{enumerate}

\usepackage{braket}
\usepackage{amsfonts}
\usepackage{amsmath}
\usepackage{amssymb}

\usepackage{color}
\usepackage{soul}

\usepackage[dvipsnames,svgnames,table]{xcolor}
\usepackage{hyperref}
\usepackage{float}
\usepackage[english]{babel}
\usepackage[caption=false]{subfig}

\usepackage{graphicx}
\hypersetup{
  breaklinks = {true},
  citecolor = {blue},
  colorlinks = {true},
  linkcolor = {red},
}
\usepackage[dvipsnames]{xcolor}

\usepackage[normalem]{ulem} 

\hypersetup{
pdfstartview={FitH},
colorlinks=true,    
linkcolor=NavyBlue, 
citecolor=Maroon,   
filecolor=NavyBlue, 
urlcolor=NavyBlue   
}

\usepackage{color}

\begin{document}
	\title{Breaking of Lorentz invariance caused by the interplay between spin-orbit interaction and transverse phonon modes in quantum wires}
\date{\today}

\author{D. V.  Efremov}  
\affiliation
{IFW Dresden, Helmholtzstr. 20, 01069 Dresden, Germany}

\author{Weyner Ccuiro}
\affiliation
{The Abdus Salam International Centre for Theoretical Physics, Strada
	Costiera 11, I-34151, Trieste, Italy}
 \affiliation
 {International School for Advanced Studies (SISSA), via Bonomea 265, I-34136 Trieste, Italy}

\author{Luis E. F. {Foa Torres}}
\affiliation
{Departamento de Fısica, Facultad de Ciencias Fısicas y Matematicas, Universidad de Chile, Santiago, 837.0415, Chile}

\author{M. N.  Kiselev}
\affiliation
{The Abdus Salam International Centre for Theoretical Physics, Strada
	Costiera 11, I-34151, Trieste, Italy}







\begin{abstract}

We investigate Lorentz invariance breaking in quantum wires due to Rashba spin-orbit interaction and transverse phonons. Using bosonization, we derive an effective action coupling electronic and mechanical degrees of freedom. Stikingly, at a quantum phase transition between straight and bent wire states, we find a gapped phonon mode and a gapless mode with quadratic dispersion, signaling the breaking of Lorentz invariance. We explore stability conditions for general potentials and propose nano-mechanical back-action as a sensitive tool for detecting this transition, with implications for Sliding Luttinger Liquids and dimensional crossover studies.
\end{abstract}

\maketitle

\textit{Introduction and overview of key findings.--} Physics shows a surprising hierarchy of alternating Lorentz and Galilean invariance, a fundamental principle governing the behavior of objects moving at relativistic speeds. As we descend to non-relativistic energies, the Lorentz invariance gives way to Galilean invariance. Intriguingly, in condensed matter physics, lower energy excitations once again obey Lorentz invariance.  A prime example is the quasiparticles in one-dimensional systems~\cite{Gogolin1999,Giamarchi2004}. Here,  electrons are described by the Luttinger liquid model~\cite{Luttinger1963,Mattis1965,Haldane1981a} whose excitations are bosons, obeying Lorentz invariance. Similarly, the mechanical degrees of freedom, described by phonons, also demonstrate Lorentz invariance.

Nano-electromechanical systems have captivated researchers for decades~\cite{Cleland2003}, offering a wide array of potential applications and research avenues. Their small size and high sensitivity make them invaluable tools in novel microelectronics \cite{Zhu2020,Hsu2023,Haroun2021}. At the cutting edge are ultrathin membranes and nanowires, fabricated from diverse materials including silicon \cite{Roberts2006,Striemer2007}, metals \cite{Jin2007}, graphene \cite{Lee2008}, carbon nanotubes \cite{Raychaudhuri2006}, and many others \cite{Hsu2023}. These systems now operate in the quantum-mechanical regime, where the behavior of the electronic component, when considered in isolation, is well understood. The most pronounced quantum-mechanical effects manifest in lower-dimensional systems, particularly nanowires. It is well established that many one-dimensional systems are accurately described by the Luttinger liquid model \cite{Kane1992,Schulz1998,Gogolin1999,Giamarchi2004}. Theoretical~\cite{Anderson1991,Wen1990,Emery2000,Sondhi2001} and experimental studies~\cite{Badawy2019,iorio_vectorial_2019} also point out Luttinger liquid physics in dimensions larger than one, with recent experiments in twisted bilayer tungsten ditelluride ($tWTe_2$)~\cite{Wu2024} which inspired further theory efforts~\cite{Yu2023}. 
However, to further advance these devices, it is crucial to comprehend the quantum effects arising from the coupling between mechanical motion and electric current.
Recent interest has also focused on the interactions between electrons and bosonic modes in cavities~\cite{Nguyen2023,Arwas2023,Nguyen2024}, expanding our understanding of these systems.

The coupling of the electrons of the nanowire in an external magnetic field with the elastic modes of the wire was first considered by Ahn and co-authors \cite{Ahn2004, Kim2008, Yi2010}. In this model, the coupling of the electric charge Luttinger liquid with the elastic modes of the nanowire occurs due to the Lorentz force as a result of the perpendicular displacements of the wire in the magnetic field. Here we show that the Rashba spin-orbit interaction couples the charge to the elastic modes of the wire via a spin current, which occurs due to transverse displacements. As 
a result of the interaction of the two corresponding Goldstone modes, which correspond to the Luttinger liquid and the elastic modes, the charge transforms into a Higgs boson and a Goldstone mode with quadratic dispersion. 
This interplay leads to intriguing physical phenomena, including the breaking of Lorentz invariance under specific conditions. In the following we present our Hamiltonian model, derive an effective action using bosonization techniques, and explore the resulting mode structure and stability conditions under various potential parameters.

\begin{figure}
	\centering
	\includegraphics[width=0.7\linewidth]{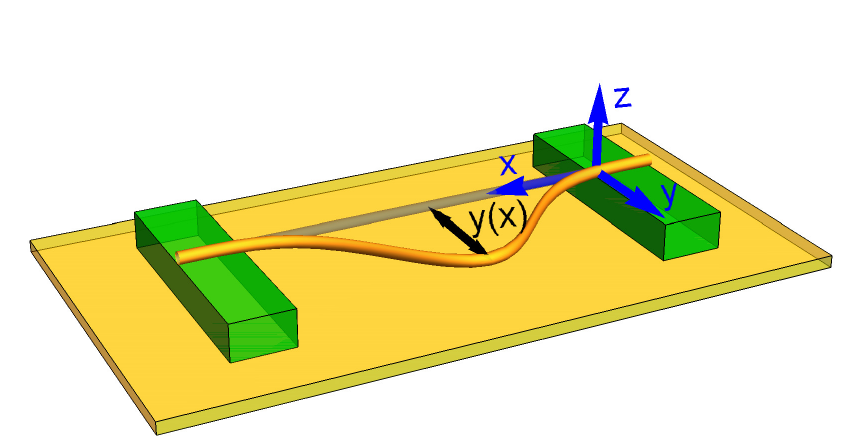}
	\caption{Schematic representation of the experimental setup. The orange beam represents a suspended double clamped nanowire with a displacement $y(x)$. Green supports can also be used as electric contacts.}
	\label{fig1}
\end{figure}
\textit{Hamiltonian model.--} The key component of the model considered here is the Rashba spin-orbit interaction arising from the external potential.  
It is described by the following Hamiltonian for a single electron:  
\begin{equation}\label{eq.1wirw.H0.1}
	\begin{split}
		\hat{H} = 
		\frac{\hat{p}^2_x}{2m} + \alpha_{so} (\hat{\vec{p}}\times \vec\sigma )\cdot \nabla U(y,z) + U(y,z),
	\end{split}
\end{equation}
where $\hat{p}_x$ is the momentum operator, 
$m$ stands for the electron mass, $\alpha_{so}$ denotes the Rashba spin-orbit coupling strength, $\vec\sigma$ represents the Pauli matrices which describe the electron's spin,  and $U(y,z)$ is the spatially dependent confining potential. It's important to note that the electric field $\vec{E}=-\nabla U(y,z)$, representing the gradient of the potential, is nonvanishing due to the absence of inversion symmetry in the system.

Following \cite{Santos2015} we assume  the external potential as  
\begin{equation}
	\label{eq.potential}
	U(y,z) = b_y y^2/2 + b_z z^2/2 + c y z. 
\end{equation}%
By a special choice of the potential $U(y,z)$ it is possible to bring the single-particle Hamiltonian Eq.~(\ref{eq.1wirw.H0.1}) into a quadratic form, similar to the model of $N$ electrons moving in a synthetic magnetic field and described by the Hamiltonian $h = 1/(2m)(\hat{p}_x - A_x)^2$. One should note, however, that this synthetic magnetic field is not breaking time-reversal symmetry and therefore preserves Kramers' degeneracy. In our problem, it is realized with the following conditions  
 \begin{equation} \label{eq.potential.for.synthetic.magnetic.field}
	2b_{y,z} =  (m\alpha_{so}^2)^{-1}\pm \sqrt{(m \alpha_{so}^2)^{-2} - 4c^2}.
\end{equation} 
where  $c$ is kept as a free parameter. Under this condition the Hamiltonian takes the following form:  
\begin{equation}
	\label{eq.hamiltonian.synthetic.magnetic field}
	\begin{split}
			\hat{H} = \!	\frac{1}{2m}\! \left(\hat{p}_x  + m \alpha_{so}  [(b_z z\!+\!  c y )\sigma_y \!-\! (b_y y\!+ \! c z )\sigma_z]  \right)^2.  
		\end{split}
\end{equation}

To demonstrate the basic properties of the dynamical model with a synthetic magnetic field, we choose  $b_y = b_z = c = (2m\alpha_{so}^2)^{-1}$.      
Rotation in the spin $(\sigma_y -\sigma_z )\sqrt{2} \to \sigma_{z}$, $(\sigma_y + \sigma_z )\sqrt{2} \to \sigma_{y}$ and the coordinate space $(z + y)/\sqrt{2} \to y $, $( y-z)/\sqrt{2} \to z $ simplify the equation to the form: 
\begin{equation}
	\begin{split}
		\hat{H} = 	\frac{1}{2m} \left( \hat{p}_x + \frac{1}{\alpha_{so}} \sigma_{z} y  \right)^2. 
	\end{split}
\end{equation}

With this the electronic part of the Hamiltonian can be written as 
\begin{equation}
	\begin{split}
		H_{e} =  \frac{1}{2m}\sum_{\substack{ \alpha,\beta=\pm}}&\int dx  \Psi_{\alpha}^\dagger(x)\left(\hat{p}_x + \frac{1}{\alpha_{so}}u_y\sigma^{z}_{\alpha}  \right)^2  \Psi_{\alpha}(x) 
		\\ & +\frac{1}{2}    \int dx dx' V(x-x')\rho(x)\rho(x'), \nonumber
	\end{split}
\end{equation} 
where $\alpha= \pm 1$ are the eigenstate indexes of the diagonalized Hamiltonian and $u_y(x) = y(x) - y_0$ is $y$ component of the transverse displacements from the equilibrium position and $y_0$ is the equilibrium position (see Fig.~\ref{fig1}). The second term describes the density-density interaction. 
Here $\rho(x) = \sum_{\substack{ \alpha=\pm}} \Psi_{\alpha}^\dagger(x)\Psi_{\alpha}(x)$ and $V(x-x')$ is the electron-electron interaction. We note that there are other possible consequences of spin-orbit interaction and its interplay with electron-electron interactions such as those considered in~\cite{Gritsev2005} which are beyond our present scope.

After linearization of the spectra close to the Fermi-level and the standard procedure of bosonization the Hamiltonian~\cite{Gogolin1999,Giamarchi2004} takes the following form: 
\begin{equation}\label{key}
	\begin{split}
		H_e = &
	\sum_{s=\pm}	\frac{v}{2\pi}\int dx \left( g (\partial_x  \theta_s - \alpha_{so}^{-1} \sigma_z u_y)^2 + g^{-1} (\partial_x \phi_s)^2   \right) \nonumber
	\end{split}
\end{equation}
where $v= v_F/g$ is the electronic velocity and $v_F$ is Fermi velocity. For short-range interaction $V(x-x') = V \delta(x-x')$, $g$ is given by $g = (1+V/(\pi v_F))^{-1/2}$. For simplicity we assume $g=1$. The full expression can be easily recovered.

The mechanical degrees of freedom is given by the elastic response of the wire on the transversal strain. 
We study a one-dimensional suspended double clamped beam of mass density $\rho_m$ and stiffness $\mathbf{T}$ extending along the $x$ direction. We assume that $y_0(x)=z_0(x)=0$ corresponds to equilibrium.
The mechanical part is related to the strain in the beam, which is associated with small transverse displacements along the $y$ and $z$ axes from the equilibrium position: $u_y(x) = y(x)-y_0$ and $u_z(x) = z(x)-z_0$.  The corresponding energy depends on the gradient of the displacements, and its behavior is captured by Hamiltonian:
\begin{equation}\label{eq.1wire.Hu}
	\begin{split}
		H_u = \sum_{\zeta = y,z}\int dx &\left( \frac{\pi^2_{\zeta}}{2\rho_m} + \frac{\mathbb{T}}{2}\left(  \frac{\partial u_\zeta}{\partial x} \right)^2 \right) 
	\end{split}
\end{equation}
Here 
$\pi_{\zeta}(x)$  are the conjugated momenta of $u_\zeta(x)$, $\zeta= y,z$, satisfying the commutation relations $[u_\zeta(x),\pi_{\zeta'}(x')] = i \delta(x-x') \delta_{\zeta\zeta'}$  .    

\textit{Effective action.--} Using the canonical transformation we get the action from the Hamiltonian. It takes the form:    
\begin{equation}\label{eq.nanowire.action.Lutitnger}
	\begin{split}
&S = S_{LL}+S_{u}+ S_{LL-u} \\
	&S_{LL}	 = \frac{v}{2 \pi g} \!\sum_{\alpha=\pm} \int_0^\beta d\tau \int_{-\infty}^\infty \!\!\!dx \left(   v^{-2}(\partial_\tau \phi_\alpha )^2  \right.  \left. + (\partial_x \phi_\alpha)^2  \right) \\ 
		&S_u =  \frac{\mathbb{T}}{2 }\sum_{\eta = y,z} \int_{0}^{\beta}d\tau \int_{-\infty}^{\infty} dx\left(  v_s^{-2}(\partial_\tau u_\eta )^2+ (\partial_x u_\eta)^2 \right)  \\
		&S_{LL-u}=- \frac{i \alpha_{so}}{\pi}^{-1}\! \!\!\int_{0}^{\beta}\!\!d\tau\!\! \int_{-\infty}^{\infty}\!\! dx ( \partial_\tau\phi_+ -\partial_{\tau}\phi_-)  u_y.   
	\end{split}
\end{equation}
$v_s$ is the phonon sound velocity and $\beta = 1/T$.
Here we would like to note that the last term corresponds to the coupling of the transverse displacement to the spinor current 
$
J  = \frac{e}{mc \alpha_{so}} (\partial_t \phi_+ - \partial_t \phi_- ) 
$, which emerges due to deviation of the beam from the equilibrium. 
It is worth to introduce new fields $\phi_c$ + $\phi_s$: 
\begin{eqnarray*}
	 \phi_c = \frac{1}{\sqrt 2} (\phi_++\phi_-),\;\;\;\;\;
	 \phi_s = \frac{1}{\sqrt 2} (\phi_+-\phi_-)
\end{eqnarray*} 
which describe the charge and spin degrees of freedom respectively.

Using the Fourier transformations of $\phi_{\alpha}(x,\tau)$ and $u_{\zeta}(x,\tau)$
\begin{eqnarray}
	\phi_{\alpha} (x,\tau) &=& \frac{1}{\sqrt{\beta L}} \sum_{k,n} e^{ikx-i\omega_{n}\tau}\phi_{kn\alpha} \\
	u_{\zeta}(x,\tau) &=& \frac{1}{\sqrt{\beta L}} \sum_{k,n} e^{ikx-i\omega_{n}\tau}u_{kn\zeta} 
\end{eqnarray}
the  Euclidean action can be written as
\begin{eqnarray}
	&&S = \sum_{k} \sum_{\omega_n}\sum_{\alpha = c,s} \frac{v}{2 \pi g } \left(\frac{\omega_{n}^{2}}{v^{2}} +k^{2}\right)\phi^{*}_{nk,\alpha}\phi_{nk,\alpha}  \nonumber \\ 
	&& +\sum_{k} \sum_{\omega_n}\sum_{\zeta = y,z} \frac{\mathbb{T}}{2} \left(\frac{\omega_{n}^{2}}{v^{2}_{s}}+k^{2}\right)u^{*}_{nk,\zeta}u_{nk,\zeta}  
\\ &&	 + \sum_{k} \sum_{\omega_n}\sum_{\zeta = y,z} \frac{\sqrt{2} \alpha_{so}^{-1}\omega_{n}}{2\pi} \big( u_{nk,y}\phi^{*}_{nk,s}
-u^*_{nk,y}\phi_{nk,s} \big) \nonumber
\end{eqnarray}

\begin{figure}
	\centering
	\includegraphics[width=0.7\linewidth]{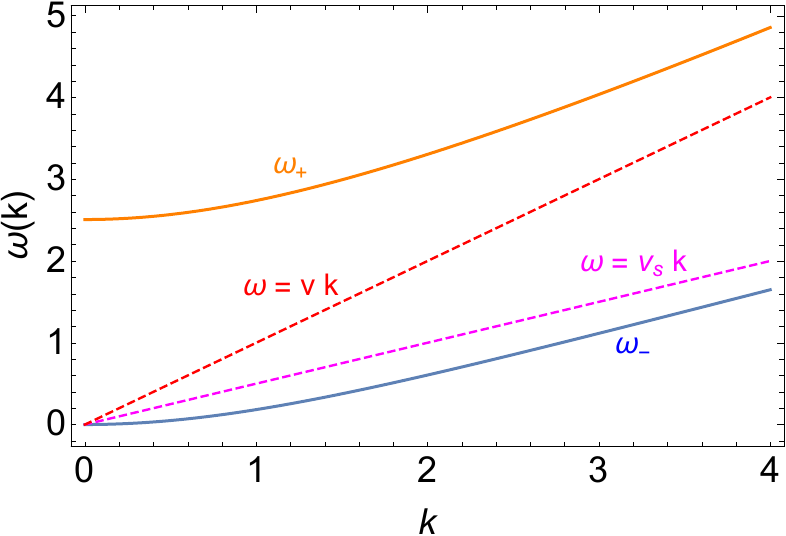}
	\caption{Eigenmodes $\omega_\pm(k)$ (solid lines).  The dashed lines denote the bosonic mode for electron $\omega = v k$, phonon $\omega = v_s k$.}
	\label{fig2}
\end{figure}

Within this action we have four modes. Two modes are decoupled. The decoupled modes are the transverse phonon mode along $z$ - direction with dispersion $\omega = v_s k$ and the charge mode of the Luttinger liquid with dispersion $\omega = vk$. The spin mode of the Luttinger liquid couples with the transverse phonon mode along the $y$ direction. It gives a gapped transverse phonon 
mode with dispersion $\omega^2= \frac{1}{2}\left((v^2_s+v^2)k^2+\Lambda^2 \right) + \sqrt{\left((v^2_s+v^2)k^2+\Lambda^2 \right)^2 - v_s^2 v^2 k^4}$, where $\Lambda = 2\pi v_F\rho \alpha_{so}^{-2}$ and a gapless mode 
$\omega^2 = \frac{1}{2}\left((v^2_s+v^2)k^2+\Lambda^2 \right) - \sqrt{\left((v^2_s+v^2)k^2+\Lambda^2 \right)^2 - v_s^2 v^2 k^4}$. The gapless mode reflects the breaking of the Lorentz invariance of the bosonic mode of the Luttinger liquid.  Instead of the linear dispersion, which corresponds to the Lorentz invariance of the Luttinger liquid, the dispersion is $\omega \sim k^2$, which corresponds to the Galilean invariance. The spectrum of the obtained quasiparticles are shown in Fig. 2

\textit{General case.--} In the previous section we examined an external potential subjected to the constraint Eq.~(\ref{eq.potential.for.synthetic.magnetic.field}). Now we explore the effect the case without the constraint. 
 \begin{equation}
 	\label{eq.hamiltonian.genralized}
 	\begin{split}
 		\hat{H} = \!	\frac{1}{2m}\! \left(\hat{p}_x  + m \alpha_{so}  [(b_z z\!+\!  c y )\sigma_y \!-\! (b_y y\!+ \! c z )\sigma_z]  \right)^2  + \tilde{U}(y,z)  \nonumber
 	\end{split}
 \end{equation} 
where 
\begin{equation}
	\tilde{U}(y,z) = U(y,z) -\frac{m \alpha_{so}^2}{2}  [(b_y y+  c z )^2+ (b_z z+  c y )^2 ]
\end{equation}
Negative $\tilde{U}(y,z)$ leads to the shift of the position of the equilibrium of the beam, $u_{y,z}(0)\neq 0$. A simple calculation lead to the following conditions for stability of the system: 
\begin{equation}
	\begin{split}
	b_y - m\alpha^2_{so}(b_y^2 +c^2) > 0,\;\;\;\;
	b_z - m\alpha^2_{so}(b_z^2 +c^2) > 0 \\
	(c(1-m\alpha_{so}^2 (b_y+b_z)))^2< (b_z - m\alpha^2_{so}(b_z^2 +c^2))
	\\ \times (b_y - m\alpha^2_{so}(b_y^2 +c^2))
	\end{split}
\end{equation}

\begin{figure}
	\centering
\includegraphics[width=0.6\linewidth]{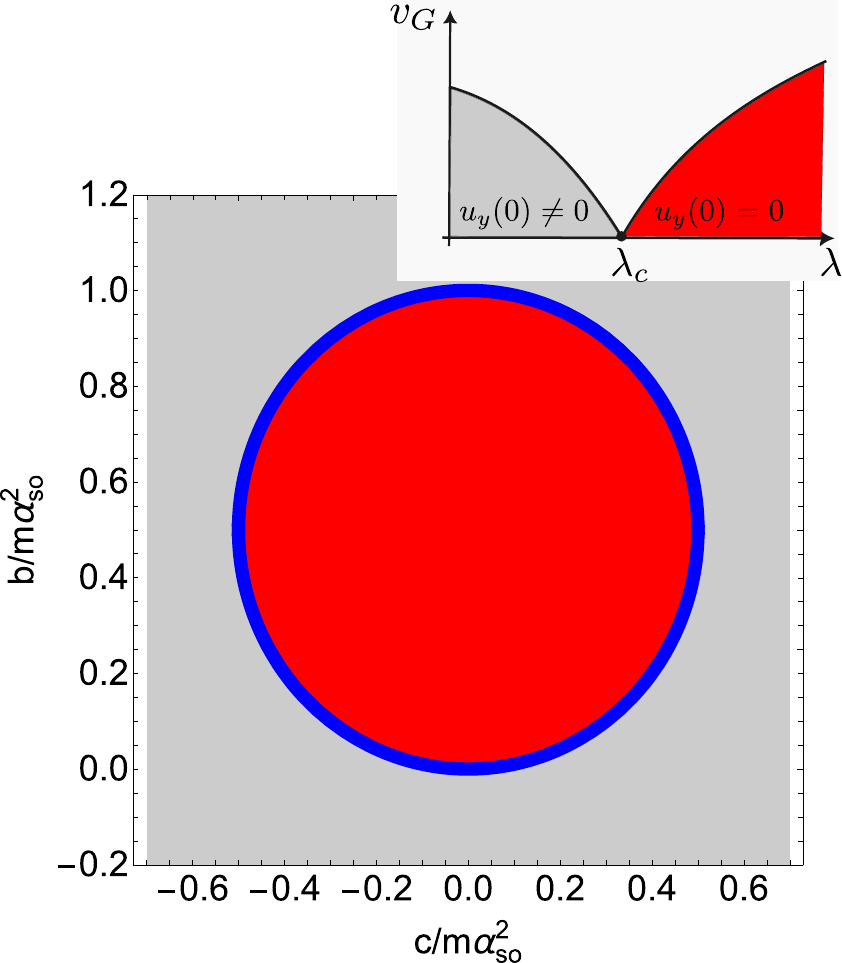}
	\caption{Schematic phase diagram of possible coefficients $b= b_{z,y}$ vs $c$. The critical region is in blue. The outside grey area corresponds to the region of instability.  The inset represents the same transition as a function of $\lambda$.}
	\label{fig3}
\end{figure}

We introduce the small transverse displacements 
as $z-z_0 = r \cos \varphi$ and  $y-y_0 = r \sin \varphi$.
Rotation in spin space we get 
   \begin{equation}
  	\label{eq.hamiltonian.genralized}
  	\begin{split}
  		\hat{H} = \!	\frac{1}{2m}\! &\left(\hat{p}_x  + \Lambda r \sigma_z]  \right)^2  
  		+\tilde{U}
  	\end{split}
  \end{equation} 
with 
$$\frac{\Lambda(\varphi)}{(m \alpha_{so})} =\left[\!(b_y \sin\varphi+  c \cos\varphi )^2 \!+ (b_z \cos\varphi+  c \sin\varphi )^2 \right]^{1/2}. $$
The effective potential is 
\begin{equation}
	\label{eq.effective.external.potential}
\begin{split}
\tilde{U}&=\!\!  \frac{r^2}{2}(b_y \sin^2\varphi+b_z \cos^2\varphi+c \sin(2\varphi)) \\ 
&-\frac{ r^2 \Lambda^2(\varphi) }{2m} = r^2 \frac{K(\varphi)}{2} 
\end{split}
\end{equation}
with 
\begin{equation}
	\begin{split}
		K(\varphi) =  \frac{\tilde{b}_z +\tilde{b}_y }{2} + \left[\left(\frac{\tilde{b}_z -\tilde{b}_y }{2}\right)^2+\tilde{c}^2\right]^{1/2} \!\!\!\! & \cos(2\varphi-\varphi_0) \\ &-c^2 m\alpha_{so}^2, 
	\end{split}
\end{equation}
where $\tan \varphi_0 = \frac{2 \tilde{c}}{\tilde{b}_z - \tilde{b}_y}$
and 
\begin{equation}
	\begin{split}
		\tilde{b}_z &= b_z (1 - m \alpha_{so}^2 b_z),\;\;\;\;\;\;
		\tilde{b}_y = b_y (1 - m \alpha_{so}^2 b_y) \\
		\tilde{c}_z &= c(1 - m \alpha_{so}^2 (b_z+b_y)) \\
	\end{split}
\end{equation}
We choose the new axes along the direction $2\varphi = \varphi_0$ and $2\varphi = \varphi_0+\pi$ and $y'$ and 
$z'$. The minimum of the effective potential corresponds to the direction of $z'$ -axis. 
The effective external potential Eq.(\ref{eq.effective.external.potential}) depends on the transverse displacements.  
It leads to the renormalization of the action of the mechanical degrees of the beam. If $K(\varphi_0/2+\pi/2)=0$ is the quantum critical point (see Fig.~\ref{fig3}). The condition Eq. (\ref{eq.potential.for.synthetic.magnetic.field}) corresponds to   $K(\varphi_0/2+\pi/2)<0$. For $K(\varphi_0/2+\pi/2)>0$ the vacuum is trivial while for  $K(\varphi_0/2+\pi/2)<0$ one needs to rederive the effective action with  non-trivial vacuum $u_{y_0}\neq 0$. 
\begin{equation}
	\begin{split}
		S_u =&  \frac{\mathbb{T}}{2 }\sum_{\zeta = y',z'}\int_{0}^{\beta}d\tau \int_{-\infty}^{\infty} dx
		\\ \times& \left(  v_s^{-2}(\partial_\tau u_{z'} )^2+ (\partial_x u_{z'})^2 + \frac{K(\varphi_\zeta)}{\mathbb{T}}u_{z'}^2 
\right),
	\end{split}
\end{equation}
  where $\varphi_{y'} = \varphi_0/2$ and $\varphi_{z'} = \varphi_0/2+\pi/2$.

Using the standard procedure we come to the action 
\begin{eqnarray}
\begin{split}
	& S = \sum_{k} \sum_{\omega_n}\sum_{\alpha = c,s} \frac{v}{2\pi g} \left(\frac{\omega_{n}^{2}}{v^{2}} +k^{2}\right)\phi^{*}_{nk,\alpha}\phi_{nk,\alpha}   \\ 
	& +\sum_{k} \sum_{\omega_n}\sum_{\zeta = y',z'} \frac{\mathbb{T}}{2} \left(\frac{\omega_{n}^{2}}{v^{2}_{s}}+k^{2}+ \frac{K(\varphi_\zeta)}{\mathbb{T}}\right)u^{*}_{nk,\zeta}u_{nk,\zeta}  \\
	&	 + \sum_{k} \sum_{\omega_n}\sum_{\zeta = y',z'} \frac{\Lambda(\varphi_\zeta)\omega_{n}}{2\pi} \big( u_{nk,\zeta}\phi^{*}_{nk,s} - h.c
	\big) .  
 \end{split}
\end{eqnarray}

The potential strongly renormalizes the phonon modes. Negative $K(\varphi)$ is attributed to a non-trivial vacuum of the system while the vacuum for positive $K(\varphi)$ is trivial. In both cases, away from the quantum critical point, Lorentz invariance is preserved and the spectrum of the lowest energy Goldstone mode $\omega\propto k$. The Lorentz invariance is broken at the quantum critical point resulting Galilean invariant Goldstone mode being $\omega\propto k^2$.

\paragraph{Correlation functions}
To see the consequences of the coupling the electronic and mechanical degrees of freedom, we calculate the 
correlation functions $\langle u_z(q,\omega_n) u_z(q,\omega_n) \rangle$ and $\langle \phi_c(q,\omega_n) \phi_c(q,\omega_n) \rangle$. To focus on the most important features of the coupling assuming that the stiffness of the mechanical degrees of freedom is strongly anisotropic, leading to the survival of only one of the mechanical modes. This assumption results in a linear polarisation of the mechanical motion. The general case will be considered elsewhere.   

Using the standard procedure of integration of complimentary degrees of freedom we get
\begin{equation}
\begin{split}
	\langle u^*_{k\omega_{n}\xi} u_{k\omega_{n}\xi} \rangle =  
	\frac{1}{\rho_m} \frac{(\omega^2_{n} + v^2 k^2)}{D}     
\end{split}
\end{equation}
and 
\begin{equation}
	\langle \phi^*_{k\omega_{n},s} \phi_{k,\omega_{n},s} \rangle = 
	\frac{\pi v_F (\omega^2_{n} + v_s^2 k^2+K/\rho_m)}{D}
\end{equation}
where 
\begin{equation}
    D = ((\omega^2_{n} + v^2 k^2)(\omega^2_{n} + v_s^2 k^2+K/\rho_m)+\tilde{\omega}^2 \omega_n^2)
\end{equation}
and $\tilde{\omega}^2 = \Lambda^2 v_F/\pi\rho_m$.
\begin{figure*}
    \centering
\includegraphics[width=0.95\textwidth]{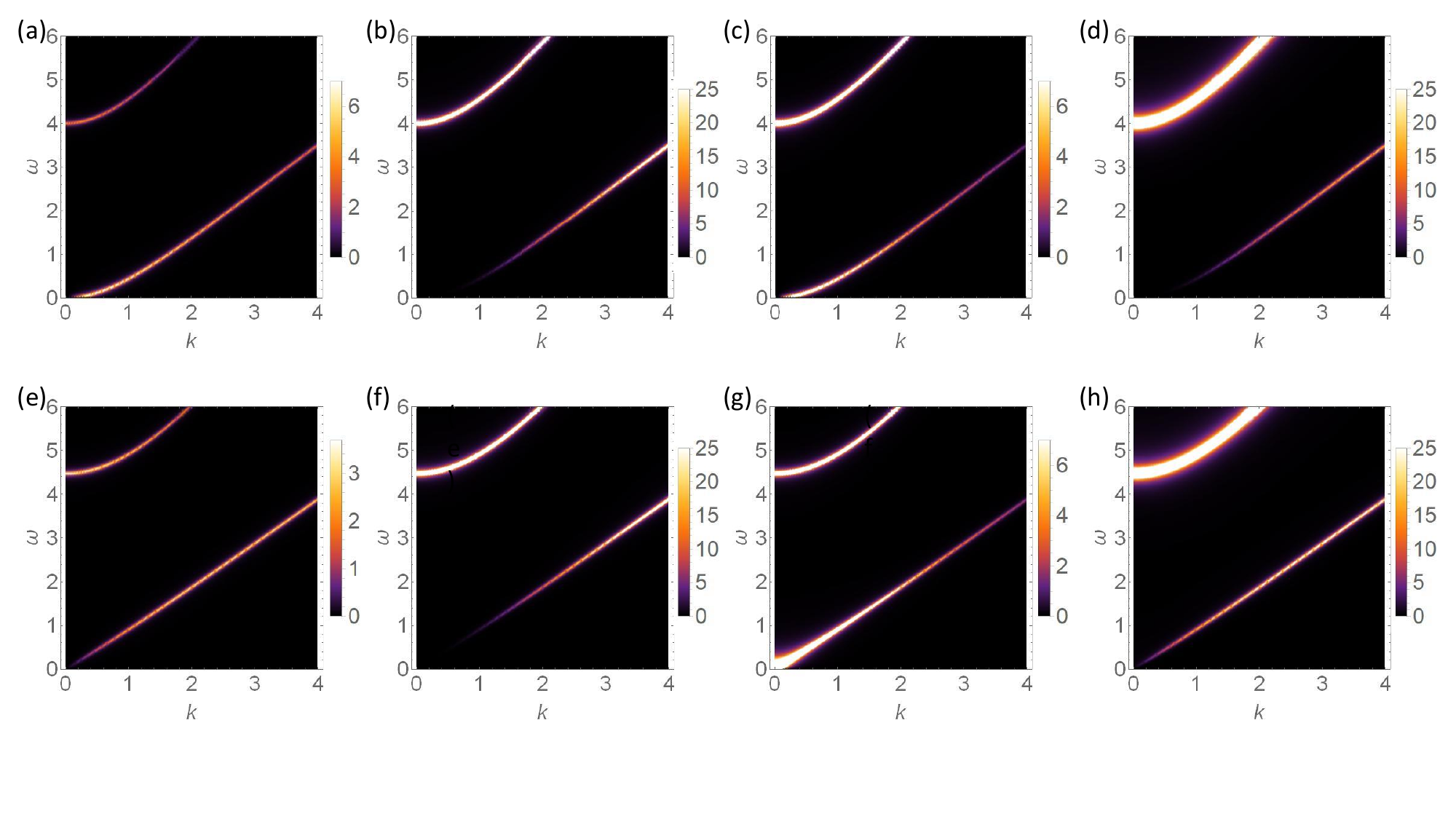}
\caption{Correlations functions a) $Im\langle u^*_y(k,\omega) u_y(k,\omega)\rangle $; b) $\omega^2Im \langle u^*_y(k,\omega) u_y(k,\omega)\rangle $; c) $Im\langle \phi^*_s(k,\omega) \phi_s(k,\omega)\rangle $; d) $\omega^2 Im\langle \phi^*_s(k,\omega) \phi_s(k,\omega)\rangle $  at the the position of the quantum critical point. Plots e) - h) show the same correlations functions as in a)-c) away from the quantum critical point (see the phase diagram on Fig. \ref{fig3}). The parameters are  $v_s = 1$, $v_F = 2$,
$\rho = 1$, $\Lambda = 1$. Parameter $K=0$ for a) - d) and $K=4$ for e) - h).
}    
    \label{fig4}
\end{figure*}
Performing analytical continuation we get the correspondent response functions. The imaginary parts of the response functions are shown in Fig. 4. 

\textit{Mechanical back action as a tool for quantum spectroscopy of the Luttinger liquids.}
One sees by comparing plots of mechanical correlation functions on Fig. \ref{fig4}  at (a), b) and away (e), f) the QCP that the character of the Goldstone mode spectrum (quadratic vs linear) and the re-distribution of the low frequency spectral weight allows to identify the position of the QPT (see Fig. \ref{fig3} insert) with a high accuracy even without measuring the quantum transport correlation functions (c),g) and (d),h)  which provide complementary information about corresponding Goldstone spectra and the spectral weights. The mechanical back action information is important as the Rashba spin-orbit interaction can be responsible for opening a gap in the charge bosonic correlation function which in turn will influence the elastic Goldstone modes. Yet another interesting questions related to behaviour of either driven or excited mechanical system are associated with effects of controllable non-linearity of elastic modes (interaction effects), possibility of parametric amplification and dissipative dynamics associated with mechanical quality factor (friction).

\textit{Final remarks and conclusions.--} The solution of a single-wire model represents a  very important step in understanding of the behavior of array of quantum wires coupled either by tunneling or by interaction (mechanically driven Sliding Luttinger Liquids (SLL). We summarize below the key findings, theoretical and experimental challenges
and relevance of the prediction to the state-of-art nano-mechanical and quantum transport experiments.

\textit{Possible experimental realization.} The $InSb$ and $InAs$ high mobility nanowires~\cite{Badawy2019,iorio_vectorial_2019} are some favorable materials for proposed experiment. The bulk $g$-factor $|g|\sim 50$ in $InSb$ is a hallmark for strong Rashba spin-orbit interaction. The "state-of-art" for $InSb$ fabrication~\cite{Badawy2019} is $\sim 10$-microns long wires of 80-100 nm diameter. The large bulk electron's mobility of $7.7\times 10^4\;{\rm cm}^2/VS$ at $T=300 K$ and large Rashba parameter $\alpha_R=0.45-0.64 eV \textup{~\AA}$ makes it a very promising material for the nano-mechanical setup. While the 
$g$-factor of bulk $InAs$ $|g|\sim 10$ and typical mobility
$\sim 1-2\times 10^4\;{\rm cm}^2/VS$~\cite{iorio_vectorial_2019} is not as high as of $InSb$, 
this material can also be used for proposed experiments.

\textit{From a single wire to few coupled quantum wires}: the next crucial step towards Sliding Luttinger Liquids (SLL) and dimensional crossover from 1D to quasi-2D quantum field theories involves solving the two-wire problem coupled by tunneling or interaction. A particularly intriguing question, both theoretically and experimentally, relates to the quantum noise associated with injecting charge or spin current in one wire and measuring the noise in the second. This effect is expected to be analogous to the quantum noise observed in co-propagating edge modes in Integer Quantum Hall (IQH) $\nu=2$ devices~\cite{Milletari2013}. By adjusting the electric circuit, one can also model counter-propagating modes. The two-wire model serves as a minimal system for studying quantum interference effects controllable by interaction.

Nano-mechanically coupled Sliding Luttinger Liquids offer a promising avenue for investigating the interplay between electronic and elastic degrees of freedom in nano-mechanically driven Integer and Fractional Topological Insulators. This model enables the study of 2D metal-insulator quantum phase transitions (QPTs) in bulk and the emergence of gapless chiral modes at the edge. The simplest model describing a transition to the Integer Topological Insulator state, analogous to the transition to the IQH regime in a synthetic magnetic field~\cite{Kane1992}, can be formulated by accounting for tunneling between neighboring wires and adjusting the chemical potential to the position of the bulk gap~\cite{Santos2015}. In the Fractional Topological Insulator state, the bulk gap opens due to inter-wire interaction effects~\cite{Santos2015}. This allows for the exploration of various bulk instabilities driven by phonons and their influence on chiral edge modes. Additionally, elastic modes associated with vibrations of a membrane in a Corbino disk geometry provide access to modeling 2D transverse phonons coupled to the chiral mode at the edge.

In summary, we have analyzed a nano-mechanical system consisting of a double-clamped suspended metallic nano-wire subject to an external dc electric field. The transverse flexural elastic modes of the wire couple to the electronic charge degrees of freedom, with the coupling strength controlled by the Rashba spin-orbit interaction in the wire. We derived the hydrodynamic action of the model and demonstrated that it is described by bosonic Gaussian theory. Our analysis reveals a quantum phase transition between two gapless phases with trivial and non-trivial vacua, corresponding to straight and bent equilibrium positions of the wire, respectively. We showed that the dispersion of the softest Goldstone mode transitions from sound-like to quadratic at the quantum phase transition point. The breaking of Lorentz invariance at the QPT is attributed to the current-displacement character of the boson-boson interaction. Our findings suggest that measuring the back-action of the nano-mechanical system provides a highly sensitive and efficient spectroscopic tool for detecting the QPT position, complementing quantum transport experiments. We propose potential experiments with suspended metallic nano-wires to verify the theoretical predictions of this model.

\textit{Acknowledgments} We thank B.L. Altshuler, Ya. Blanter, A.Chubukov, Y.Gefen, Th. Giamarchi, D. Maslov, C. Mora, F. von Oppen, S. Sachdev and J. Schmalian for fruitful discussions. The work of M.N.K is conducted within the framework of the Trieste Institute for Theoretical Quantum Technologies (TQT) and supported in part by the NSF under Grants No. NSF PHY-1748958 and No. PHY-2309135. M.N.K. acknowledges support of the IHP (UAR 839 CNRS-Sorbonne Universite)
and LabEx CARMIN (ANR-10LABX-59-01) L.E.F.F.T. acknowledges financial support by ANID FONDECYT (Chile) through grant number 1211038, The Abdus Salam International Center for Theoretical Physics and the Simons Foundation, and by the EU Horizon 2020 research and innovation program under the Marie-Sklodowska-Curie Grant Agreement No. 873028 (HYDROTRONICS Project). D.V.E. acknowledges the financial support of DFG (grant numbers 529677299, 449494427) and hospitality of Abdus Salam International Center for Theoretical Physics. 
\bibliographystyle{apsrev4-1_title}
\bibliography{sLL}
\end{document}